\documentclass[fleqn,prl,twocolumn]{revtex4}

\usepackage{amsmath,graphicx}
\usepackage{dcolumn}
\usepackage{verbatim}

\begin{document}

\title{High-precision solution of the Dirac Equation for \\
the hydrogen molecular ion by an iterative method}

\author{Hugo D. Nogueira$^1$, Vladimir I. Korobov$^2$,
and Jean-Philippe Karr$^{1,3}$}
\affiliation{$^1$Laboratoire Kastler Brossel, Sorbonne Universit\'e, CNRS, ENS-Universit\'e PSL, Coll\`ege de France, 4 place Jussieu, F-75005 Paris, France}
\affiliation{$^2$Bogoliubov Laboratory of Theoretical Physics, Joint Institute for Nuclear Research, Dubna 141980, Russia}
\affiliation{$^2$Universit\'e d'Evry-Val d'Essonne, Universit\'e Paris-Saclay, Boulevard Fran\c cois Mitterrand, F-91000 Evry, France}

\begin{abstract}
The Dirac equation for H$_2^+$ is solved numerically using an iterative method proposed by Kutzelnigg [Z. Phys. D~\textbf{11}, 15 (1989)]. The four-component wavefunction is expanded in a newly introduced kinetically balanced exponential basis set. The ground-state relativistic energy is obtained with an accuracy of $10^{-20}$, which represents an improvement by several orders of magnitude, and is shown to be in good agreement with results obtained from perturbation theory. Highly accurate relativistic wavefunctions are obtained, which is a first step towards nonperturbative calculations of the one-loop self-energy correction in hydrogen molecular ions.
\end{abstract}

\maketitle

The determination of quantum states of an electron in the field of two charged nuclei is one of the most fundamental problems of quantum chemistry. At the non-relativistic level, the two-center Schr\"odinger equation has been known for a long time to lend itself to separation of variables using spheroidal (elliptic) coordinates, and can be solved with essentially arbitrary accuracy (see, e.g.,~\cite{Peek65,Ishikawa12}). However, the relativistic Dirac equation in a two-center potential~\cite{Mueller73} poses more serious difficulties. For example, the best accuracy reported so far for the ground-state energy of the H$_2^+$ molecular ion is about $10^{-13}$~\cite{Kullie01,Tupitsyn14}.

Interest in this problem has been fueled by the perspective of testing molecular QED effects in the strong-field regime through collisions between highly charged ions, which are planned to be studied in new-generation experiments at future heavy-ion research facilities~\cite{Gumberidze09,Ma17}. Of special interest is the phenomenon of spontaneous positron emission predicted to occur when the total charge of the nuclei is larger than the critical value $Z_{\rm cr} \approx 173$~\cite{Greiner85}, experimental signatures of which are being actively sought~\cite{Maltsev19}.

Another field of applications has recently emerged in connection with the precision spectroscopy of hydrogen molecular ions. Three rovibrational transitions in HD$^+$ have been measured with relative uncertainties in the $10^{-11}$-$10^{-12}$ range~\cite{Alighanbari20,Patra20,Kortunov21}, approaching or exceeding the current precision of theoretical predictions~\cite{Korobov17,Korobov21}. Comparison between theory and experiment has led to an improved determination of the proton-electron mass ratio~\cite{Korobov21} and improved constraints on a ``fifth force'' between hadrons~\cite{Germann21}. These results, and the fact that the experimental precision may be pushed further in the future~\cite{Schiller14,Karr14}, strongly motivate further improvement of the theoretical precision. The latter is currently limited to 7-8~$10^{-12}$ by the one-loop self-energy of the bound electron, which has been calculated in the non-relativistic QED approach up to the $m\alpha(Z\alpha)^6$~order. One way to overcome this limit would be to perform a full relativistic calculation of the one-loop self-energy, i.e. without performing the expansion in $Z\alpha$, as done for the hydrogen atom~\cite{Jentschura99,Jentschura01}. A 7-8 digit precision for this quantity would improve theoretical rovibrational transition frequencies by about a factor of 2. To achieve this, the required precision in the relativistic wavefunctions is actually much higher than the aforementioned 7-8 digits, because the self-energy is a residual effect obtained after subtraction of renormalization counterterms, leading to a serious loss of precision~\cite{Jentschura01}. This brings an important motivation to solve Dirac's equation with the highest possible accuracy.

This problem may be approached in two different ways. One can attempt a direct resolution of the two-center Dirac equation, for which the most accurate results so far have been obtained by the finite-element method~\cite{Kullie01,Yang91} and by the Dirac-Fock-Sturm method~\cite{Tupitsyn14}. Alternatively, one can use a perturbative approach where the energy and wavefunction are expanded in powers of $c^{-2}$. The first-order correction for H$_2^+$ has been obtained with high numerical accuracy using the Breit-Pauli effective Hamiltonian~\cite{Tsogbayar06}. Higher-order effective Hamiltonians can also be derived using Foldy-Wouthuysen transformations~\cite{Douglas74,Pachucki05} or in the NRQED framework~\cite{Haidar20}. This has allowed the second-order ($c^{-4}$) correction to be evaluated~\cite{Korobov07}. However, it would be difficult to extend this method to higher orders, in particular due to the increasingly singular behavior of the effective operators. The direct perturbation theory (DPT)~\cite{Rutkowski87,Kutzelnigg89,Kutzelnigg96}, expressed in terms of 4-component spinors, has been shown to avoid this problem and does not require a controlled cancellation of divergences. This method has been used to calculate the third-order ($c^{-6}$) correction in H$_2^+$~\cite{Rutkowski87,Franke92}. An iterative method based on the principles of DPT was also derived in~\cite{Kutzelnigg89} and later applied to high-Z hydrogenlike ions~\cite{Franke97}.

A perturbative approach such as DPT is especially well suited for weakly relativistic systems such as H$_2^+$; moreover, the zero-order wavefunction, which is a solution of the two-center Schr\"odinger equation, can be obtained with extremely high accuracy. For the aim of extending calculations to higher orders, the iterative method of Ref.~\cite{Kutzelnigg89} is especially attractive because no tedious algebraic manipulations are required to express relativistic corrections at any order. Previous applications of this method have been performed using Gaussian basis sets~\cite{Franke92,Franke97}. In this work, we introduce a basis set of pure two-center exponentials, which have so far only been used in nonrelativistic calculations~\cite{Tsogbayar06}. A key advantage of exponential functions is that they allow to better represent the singular behaviour of the solutions in the vicinity of the nuclei.  This allows us to improve the accuracy of the relativistic energy and wavefunction of H$_2^+$ by several orders of magnitude.

The atomic unit system ($\hbar = m = e = 1$) is used throughout. In these units, the velocity of light is $c$ regarded as dimensionless and has the value $\alpha^{-1} \approx 137$. The Dirac equation can be written as:
\begin{subequations} \label{eq_dirac}
\begin{flalign}
&H_D \psi = E \psi, \;\;\; \psi = 
    \begin{pmatrix}
    \varphi \\
    \chi
    \end{pmatrix},& \\
&H_D = (\beta\!-\!I_4) c^2 + c \boldsymbol{\alpha}\mathbf{p} + V = 
    \begin{pmatrix}
      V & c \boldsymbol{\sigma}\mathbf{p} \\
      c \boldsymbol{\sigma}\mathbf{p}  & V \!-\! 2 c^2
    \end{pmatrix},&
\end{flalign}
\end{subequations}
where $\psi$ is a Dirac spinor, and $\varphi$, $\chi$ are two-components objects representing respectively the large and small components. $\beta$ and $\boldsymbol{\alpha}$ are Dirac matrices, $\boldsymbol{\sigma}$ the Pauli matrices, and $I_4$ the $4\!\times\!4$ identity matrix. The rest mass energy $c^2$ has been subtracted from the energy. Finally, $V$ is the two-center Coulomb potential given by
\begin{equation} \label{eq_V}
    V = -\frac{Z_1}{r_1} - \frac{Z_2}{r_2} \,,
\end{equation}
where $Z_1$ and $Z_2$ are the charges of the nuclei, and $r_1$, $r_2$ the distances from the electron to both nuclei. The starting idea of DPT is to perform the following metric transformation in order to obtain the nonrelativistic limit of the Dirac equation~\cite{Kutzelnigg89}:
\begin{equation}
\begin{pmatrix}
\varphi \\
\chi
\end{pmatrix} =
\begin{pmatrix}
I_2 & 0 \\
0   & c^{-1} I_2
\end{pmatrix}
\begin{pmatrix}
\varphi \\
\tilde{\chi}
\end{pmatrix}
\end{equation}
The Dirac equation can then be rewritten as
\begin{equation}
\begin{pmatrix}
V & \boldsymbol{\sigma}\mathbf{p} \\
\boldsymbol{\sigma}\mathbf{p} & -2 + \frac{V}{c^2}
\end{pmatrix}
\begin{pmatrix}
\varphi \\
\tilde{\chi}
\end{pmatrix}
= E \begin{pmatrix}
I_2 & 0 \\
0   & c^{-2} I_2
\end{pmatrix}
\begin{pmatrix}
\varphi \\
\tilde{\chi}
\end{pmatrix}
\end{equation}
Solving the second line for $\tilde{\chi}$, one obtains
\begin{equation} \label{eq_dirac1}
\tilde{\chi} = \frac{\boldsymbol{\sigma}\mathbf{p}}{2} \varphi + \frac{V-E}{2c^2} \tilde{\chi} \,,
\end{equation}
and injecting this result into the first line yields
\begin{equation} \label{eq_dirac2}
(E - H_0) \varphi = \frac{\boldsymbol{\sigma}\mathbf{p}}{2c^2} (V-E) \tilde{\chi}\,,
\end{equation}
where $H_0 = \mathbf{p}^2/2 + V$ is the Schr\"odinger Hamiltonian. Kutzelnigg~\cite{Kutzelnigg89} proposed an iterative solution based on Eqs.~(\ref{eq_dirac1}-\ref{eq_dirac2}). The first iteration step is the solution of the Schr\"odinger equation
\begin{equation} \label{eq_se}
    H_0 \varphi_0^{(1)} = E_0 \varphi_0^{(1)} \,,
\end{equation}
where $\varphi_0^{(1)}$ represents the first component of $\varphi_0$. The second component $\varphi_0^{(2)}$ is set to zero, which corresponds to taking the zero-order solution in a spin state $S_z = 1/2$. The small components are given by $\tilde{\chi}_0 = \frac{\boldsymbol{\sigma}\mathbf{p}}{2} \varphi_0$. One then iterates over Eqs. (\ref{eq_energy}-\ref{eq_chi}):
\begin{subequations}
\begin{flalign}
&E_{i+1} = E_0 +\frac{1}{c^2}\left\langle\tilde{\chi}_0|(V - E_i) |\tilde{\chi}_i\right\rangle,& \label{eq_energy} \\
&(E_{i+1}\!-\!H_0)\Delta\varphi_{i+1} = \frac{1}{2c^2} Q \, \boldsymbol{\sigma}\textbf{\text{p}}(V\!-\!E_{i+1})\tilde{\chi}_i\,,& \label{eq_phi} \\
&\tilde{\chi}_{i+1} = \frac{\boldsymbol{\sigma}\textbf{\text{p}}}{2}\varphi_{i+1} + \frac{1}{2c^2}(V-E_{i+1}) \tilde{\chi}_i\,,& \label{eq_chi}
\end{flalign}
\end{subequations}
where the subscript $i$ refers to the iteration step, $\varphi_i=\varphi_0+\Delta\varphi_i$, and $Q = 1\!-\!|\varphi_0\rangle\langle\varphi_0|$ is a projector onto a subspace orthogonal to $|\varphi_0\rangle$. Note that Eq.~(\ref{eq_energy}) can be obtained by multiplying Eq.~(\ref{eq_dirac2}) on the left by $\varphi_0^{\ast}$ followed by space integration. This method converges faster than the perturbative expansion in powers of $c^{-2}$, especially in highly relativistic (high-$Z$) systems~\cite{Franke97}.

Let us now describe our implementation of the iterative method. The large components of the wavefunction are expanded in an exponential basis set~\cite{Tsogbayar06,Korobov07}:
\begin{subequations} 
\begin{flalign}
&\varphi^{(j)} = \sum_{i=1}^N c_i^{(j)} f_i^{(j)} \,,& \label{eq_basis} \\
&f_i^{(j)} (\textbf{r}) = e^{im^{(j)}\phi} r^{|m^{(j)}|} \left( e^{-\alpha_i r_1 -\beta_i r_2} \pm e^{-\beta_i r_1 -\alpha_i r_2} \right)& \label{eq_basis_function}
\end{flalign}
\end{subequations}
where $j=1,2$ indicates the component, $\phi$ is the angle around the internuclear axis $z$, and $r$ the distance from the electron to the internuclear axis. $m^{(j)}$ is an eigenvalue of $L_z$, the projection of the orbital angular momentum on the $z$ axis. The sign in Eq.~(\ref{eq_basis_function}) is equal to $(-1)^{m^{(j)}}$ for {\textit gerade} states and $-(-1)^{m^{(j)}}$ for {\textit ungerade} states. Since the total angular momentum projection $J_z = L_z + S_z$ is a good quantum number, for the ground ($1s\sigma_g$) electronic state and $J_z = 1/2$ one has $m^{(1)}=0$ and $m^{(2)}=1$; this also applies to the small components $\tilde{\chi}^{(j)}$~\cite{Mueller76}. 

The small components of the wavefunction are expanded in the kinetically balanced basis~\cite{Stanton84} 
\begin{subequations} \label{eq_basis}
\begin{gather}
\tilde{\chi}^{(j)} = \sum_{i=1}^N d_i^{(j)} g_i^{(j)} \,, \\
\begin{pmatrix}
g_i^{(1)} \\
g_i^{(2)}
\end{pmatrix} = \frac{\boldsymbol{\sigma}\mathbf{p}}{2}
\begin{pmatrix}
f_i^{(1)} \\
f_i^{(2)}
\end{pmatrix}
\end{gather}
\end{subequations}
Kinetic balance is a key ingredient for the numerical calculations, as discussed in~\cite{Franke97}. In particular, it allows for efficient cancellation of singularities in the right-hand side of Eq.~(\ref{eq_chi})~\cite{Kutzelnigg89}.

The matrix elements appearing in Eqs.~(\ref{eq_energy}-\ref{eq_chi}) are calculated analytically using the methods described in~\cite{Tsogbayar06,Korobov07}. In particular, those of $\frac{\boldsymbol{\sigma}\textbf{\text{p}}}{2} (V-E) \frac{\boldsymbol{\sigma}\textbf{\text{p}}}{2}$, which are needed in all three equations, can be obtained from the identity
\begin{equation} \label{eq_transformation}
\begin{aligned}
 \frac{\boldsymbol{\sigma}\textbf{\text{p}}}{2} (V-E) \frac{\boldsymbol{\sigma}\textbf{\text{p}}}{2} =
      &\, \frac{1}{8} (p^2 V + V p^2) -\frac{1}{4} E p^2  \\
      & + \frac{\pi}{2} \left[Z_1 \delta(\mathbf{r}_1) + Z_2 \delta(\mathbf{r}_2)\right] + H^{SO} \,, \\
H^{SO} = & \, \left( Z_1 \frac{\left[ \mathbf{r}_1 \!\times\! \mathbf{p} \right]}{4 r_1^3} + Z_2 \frac{\left[ \mathbf{r}_2 \!\times\! \mathbf{p} \right]}{4 r_2^3} \right) \cdot \boldsymbol{\sigma}\,,
\end{aligned}
\end{equation}
where $\delta$ is the Dirac delta function, and $H^{SO}$ the spin-orbit Hamiltonian.

The basis set is constructed in the following way. It consists of several subsets, each subset being defined by a pair of intervals in which the exponents $\alpha_i$, $\beta_i$ in Eq.~(\ref{eq_basis_function}) are generated in a pseudorandom way~\cite{Tsogbayar06,Korobov07}. The subsets are separated into two groups, see Table~\ref{tab_basisset} for an illustrative example: a ``regular'' part made of two or three intervals (depending on the internuclear distance $R$) containing small exponents (typically $\alpha_i,\beta_i<10$), and a ``singular'' part made of five or six intervals (also depending on $R$) containing large exponents (up to $10^7$). The latter part is required to accurately represent the singular behavior of the Dirac wavefunction in the vicinity of the Coulomb centers. The parameters of the basis (the interval bounds for each subset, and the relative sizes of the subsets) can be optimized by varying each parameter and selecting the values that provide the fastest convergence as a function of the basis size. In view of the large number of parameters, only a coarse optimization has been performed.

Numerical calculations are performed in octuple precision arithmetic. Unless otherwise noted, the CODATA-2018 value of the fine-structure constant, i.e. $c=137.035\,999\,084$, is used~\cite{Tiesinga21}. The convergence of our results for the equilibrium internuclear distance $R = 2$~a.u. is shown in Table~\ref{tab_r2}, and in more detail in Table~\ref{tab_r2_iterations} where energies obtained after the first four iterations are shown. Results are much more sensitive to the size of the regular basis, whereas adding more functions the singular basis results in negligibly small changes in the energy; this is why the convergence is analyzed by varying the size of the regular basis, $N_{\rm reg}$, while leaving the singular basis unchanged. Inspection of Table~\ref{tab_r2_iterations} shows that the precision is progressively degraded as the iteration order increases. Results of the fifth iteration (and beyond) are not converged; the corresponding energy correction is smaller than $10^{-22}$~a.u. and thus insignificant with respect to the achieved precision of $1 \times 10^{-20}$~a.u on the Dirac energy. The precision is mainly limited by the second iteration and to a lesser extent by the third iteration. It could in principle be improved by increasing the basis size beyond $N_{\rm reg}=300$, but this results in numerical instabilities in the resolution of the linear system in Eq.~(\ref{eq_chi}). These instabilities are likely to be linked to the improper behavior of basis functions in the vicinity of the nuclei in the case of the function $\tilde{\chi}^{(2)}$. Indeed, the kinetic balance relationship, Eq.~(\ref{eq_basis}), yields basis functions that have a finite value at the nuclei, whereas Dirac solutions for $m=1$ ($\pi$) components tend to zero.

\begin{table}[t]
\begin{center}
\begin{tabular}{@{\hspace{4mm}}c@{\hspace{4mm}}c@{\hspace{7mm}}c@{\hspace{4mm}}c@{\hspace{7mm}}c@{\hspace{3mm}}}
\hline\hline
\vrule width0pt height10pt depth4pt
$A_1$ & $A_2$  & $B_1$ & $B_2$ & $n_i$ \\
\hline
\vrule width0pt height10pt depth4pt
$0.0$              & $1.5$               & $0.0$   & $0.4$     & $100$ \\
$1.0$              & $6.0$               & $0.2$   & $2.0$     & $100$ \\
$2.0$              & $10.0$              & $0.0$   & $2.0$     & $100$ \\
\hline
\vrule width0pt height10pt depth4pt
$1.0\!\times\!10^1$    & $3.0\!\times\!10^1$     & $0.0$   & $2.0$     & $41$  \\
$3.0\!\times\!10^1$    & $3.0\!\times\!10^2$     & $0.0$   & $2.0$     & $38$  \\
$3.0\!\times\!10^2$    & $4.0\!\times\!10^3$     & $0.0$   & $2.0$     & $34$  \\
$4.0\!\times\!10^3$    & $6.0\!\times\!10^4$     & $0.0$   & $2.0$     & $31$  \\
$6.0\!\times\!10^4$    & $8.0\!\times\!10^5$     & $0.0$   & $2.0$     & $28$  \\
$8.0\!\times\!10^5$    & $1.0\!\times\!10^7$     & $0.0$   & $2.0$     & $26$  \\
\hline
\hline
\end{tabular}
\end{center}
\caption{Basis set used in numerical calculations for $R=2.0$. $[A_1, A_2]$ ($[B_1,B_2]$) are intervals in which the exponents $\alpha_i$ ($\beta_i$) are generated. $n_i$ is the number of basis functions in each subset. The total basis size is $N=498$.}\label{tab_basisset}
\end{table}

\begin{table}[t]
\begin{center}
\begin{tabular}{@{\hspace{2mm}}c@{\hspace{5mm}}c@{\hspace{5mm}}}
\hline\hline
\vrule width0pt height10pt depth4pt
$N_{\rm reg}$ & $E$ \\
\hline
\vrule width0pt height10pt depth4pt
$225$      & $\mathbf{-1.102\,641\,581\,032\,577\,164\,1}39\,937$  \\
$240$      & $\mathbf{-1.102\,641\,581\,032\,577\,164\,1}33\,856$  \\
$255$      & $\mathbf{-1.102\,641\,581\,032\,577\,164\,1}32\,196$  \\
$270$      & $\mathbf{-1.102\,641\,581\,032\,577\,164\,1}31\,380$  \\
$285$     & $\mathbf{-1.102\,641\,581\,032\,577\,164\,12}7\,416$  \\
$300$     & $\mathbf{-1.102\,641\,581\,032\,577\,164\,12}6\,607$  \\
\hline
\vrule width0pt height10pt depth4pt
extrap. & $\mathbf{-1.102\,641\,581\,032\,577\,164\,12}(1)$ \\
\hline
\hline
\end{tabular}
\end{center}
\caption{Energy of the $1s\sigma_g$ ground state for $R=2.0$ obtained using the iterative method, as a function of total size $N_{\rm reg} = n_1 + n_2 + n_3$ of the regular basis (i.e. the first three subsets in Table~\ref{tab_basisset}). The sizes of the regular subsets are $n_1 = n_2 = n_3 = N_{\rm reg}/3$. The singular part of the basis is the same as shown in Table~\ref{tab_basisset}. Bold figures are converged.}\label{tab_r2}
\end{table}

\begin{table*}[t]
\begin{center}
\begin{tabular}{@{\hspace{2mm}}c@{\hspace{3mm}}c@{\hspace{3mm}}c@{\hspace{3mm}}c@{\hspace{3mm}}c@{\hspace{3mm}}c@{\hspace{2mm}}}
\hline\hline
\vrule width0pt height10pt depth4pt
$N_{\rm reg}$ & $(E_1\!-\!E_0)\times$10$^6$ & $(E_2\!-\!E_1)\times$10$^{10}$ & $(E_3\!-\!E_2)\times$10$^{15}$ & $(E_4\!-\!E_3)\times$10$^{19}$ \\
\hline
\vrule width0pt height10pt depth4pt
$225$      &  $\mathbf{-7.366\,419\,298\,336\,650\,496\,815\,9}14\,28$ & $\mathbf{-1.183\,246\,223}\,379$ & $\mathbf{-7.743\,32}1\,61$ & $\mathbf{-3.208\,4}76$ \\
$240$      &  $\mathbf{-7.366\,419\,298\,336\,650\,496\,815\,908}\,66$ & $\mathbf{-1.183\,246\,223}\,317$ & $\mathbf{-7.743\,32}1\,72$ & $\mathbf{-3.208\,4}61$ \\
$255$      &  $\mathbf{-7.366\,419\,298\,336\,650\,496\,815\,908}\,88$ & $\mathbf{-1.183\,246\,223\,2}99$ & $\mathbf{-7.743\,32}1\,90$ & $\mathbf{-3.208\,4}36$ \\
$270$      &  $\mathbf{-7.366\,419\,298\,336\,650\,496\,815\,908}\,64$ & $\mathbf{-1.183\,246\,223\,2}90$ & $\mathbf{-7.743\,32}1\,94$ & $\mathbf{-3.208\,4}31$ \\
$285$      &  $\mathbf{-7.366\,419\,298\,336\,650\,496\,815\,908\,5}6$ & $\mathbf{-1.183\,246\,223\,2}49$ & $\mathbf{-7.743\,32}2\,13$ & $\mathbf{-3.208\,4}04$ \\
$300$     &  $\mathbf{-7.366\,419\,298\,336\,650\,496\,815\,908\,5}8$ & $\mathbf{-1.183\,246\,223\,2}42$ & $\mathbf{-7.743\,32}1\,97$ & $\mathbf{-3.208\,4}24$ \\
\hline
\hline
\end{tabular}
\end{center}
\caption{Corrections to the ground-state energy during the first four iterations for the results shown in Table~\ref{tab_r2}. Bold figures are converged. The zero-order (nonrelativistic) energy $E_0$ (not shown here) is converged to more than thirty digits, and its value can be found in Table~\ref{tab_r2_pt2}. }\label{tab_r2_iterations}
\end{table*}

\begin{table}[t]
\begin{center}
\begin{tabular}{@{\hspace{3mm}}c@{\hspace{3mm}}l@{\hspace{2mm}}c@{\hspace{3mm}}}
\hline\hline
\multicolumn{2}{c}{Dirac energy} & Ref. \vrule width0pt height10pt depth4pt\\
\hline
\vrule width0pt height10pt depth4pt
$E$ & $-1.102\,641\,581\,033\,607\,580\,05(1)$ & this work \\
    & $-1.102\,641\,581\,033\,58$              &~\cite{Kullie01} \\
    & $-1.102\,641\,581\,033\,0$               &~\cite{Tupitsyn14} \\
\hline\hline
\multicolumn{2}{c}{Direct Perturbation Theory} & Ref. \vrule width0pt height10pt depth4pt \\
\hline
\vrule width0pt height10pt depth4pt
$E_0$         & $-1.102\,634\,214\,494\,946\,461\,508\,968\,945\,318$ & this work \\
              & $-1.102\,634\,214\,494\,946\,461\,50$                 & \cite{Korobov07} \\
              & $-1.102\,634\,214\,494\,946\,462$                     & \cite{Ishikawa12} \\
$E_1^p$       & $-0.138\,332\,993\,867\,979\,584\,653\,9 $            & this work \\
              & $-0.138\,332\,993\,9$                                 & \cite{Korobov07} \\
              & $-0.138\,332\,984\,8$                                 & \cite{Franke92} \\           
$E_2^p$       & $-0.041\,727\,900\,54(1)$                             & this work \\
              & $-0.041\,711$                                         & \cite{Korobov07} \\
              & $-0.041\,727\,79$                                     & \cite{Franke92} \\
$E_3^p$       & $-0.028\,318\,426\,48$                                & this work \\
              & $-0.028\,32$                                          & \cite{Franke92} \\
              & $-0.028(2)$                                           & \cite{Rutkowski87} \\
\hline
\vrule width0pt height10pt depth4pt 
$E^p$ & $-1.102\,641\,581\,033\,607\,579\,88$                 & this work \\
\hline
\hline
\end{tabular}
\end{center}
\caption{Comparison of the Dirac energy obtained in this work for the ground state at $R=2.0$ with previous results, and with results from DPT. The same value of $c$ as in earlier works~\cite{Kullie01,Tupitsyn14},  $c=137.035\,989\,5$, has been used. Energy corrections $E_i^p$ are given in units of $c^{-2i}E_h$, where $E_h$ is the Hartree energy. An estimate of the Dirac energy from DPT is obtained as $E^p = E_0 + \sum_{i=1}^3 c^{-2i} E_i^p $. In the results of 'this work', all digits are converged unless otherwise noted.}\label{tab_r2_pt2}
\end{table}

As a cross-check of our results, we have also implemented DPT up to third order in the same basis set. To the best of our knowledge, no finite expression for the fourth-order correction has been obtained so far. Energy corrections at successive orders in $c^{-2}$ are expressed as~\cite{Rutkowski87,Kutzelnigg89}
\begin{equation}\label{eq_e_dpt}
\begin{aligned}
E_1^p = &\,\left\langle\tilde{\chi}_0|(V - E_0)|\tilde{\chi}_0\right\rangle \,, \\  
E_2^p = &\,\left\langle\tilde{\chi}_0|(V - E_0)|\tilde{\chi}_1^p\right\rangle - E_1^p \left\langle\tilde{\chi}_0|\tilde{\chi}_0\right\rangle \,, \\
E_3^p = &\,\left\langle\tilde{\chi}_1^p|(V - E_0)|\tilde{\chi}_1^p\right\rangle  
 - E_1^p \left\{\left\langle\tilde{\chi}_0|\tilde{\chi}_1^p\right\rangle \right. \\
 &+ \left. \left\langle\tilde{\chi}_1^p|\tilde{\chi}_0\right\rangle + \left\langle\varphi_1^p|\varphi_1^p\right\rangle \right\} - E_2^p \left\langle\tilde{\chi}_0|\tilde{\chi}_0\right\rangle \,,
\end{aligned}
\end{equation}
\begin{subequations}
where the first-order perturbation wavefunctions $\varphi_1^p$ and $\tilde{\chi}_1^p$ are given by
\begin{align}
(H_0\!-\!E_0)\varphi_1^p &= -\frac{\boldsymbol{\sigma}\textbf{\text{p}}}{2}(V\!-\!E_0)\tilde{\chi}_0\,,& \label{eq_phi2} \\
\tilde{\chi}_1^p &= \frac{\boldsymbol{\sigma}\textbf{\text{p}}}{2}\varphi_1^p + \frac{1}{2}(V-E_0) \tilde{\chi}_0\,.& \label{eq_chi2} 
\end{align}
\end{subequations}
For DPT calculations, we varied the basis size up to $N_{\rm reg}=240$; it was not useful to increase it further because the precision of the Dirac energy value obtained from DPT is limited by the unevaluated fourth-order correction. Table~\ref{tab_r2_pt2} shows a summary of our results and comparison with previous works. Satisfactory agreement is obtained~\cite{DPT-comment}, and the precision is improved by several orders of magnitude both for the Dirac energy and for DPT results. The difference between results obtained from the iterative method and from DPT amounts to $1.6\!\times\!10^{-19}$~a.u., which is consistent with the expected magnitude of the fourth-order correction. From this difference one may deduce the estimate $E_4^p \sim -0.020 \; c^{-8} E_h$.

Finally, we have applied the iterative method for other values of the internuclear distance $R$. Results are shown in Table~\ref{tab_allrR}. The general behavior of the method is similar, with the first four iterations being well-converged, but the achieved precision is higher at small ($R \leq 1$~a.u.) and large ($R \geq 5$~a.u.) internuclear distances. This observation supports the hypothesis that the precision is limited by imperfect representation of the component $\chi^{(2)}$. Indeed, the $m=1$ ($\pi$) components appear as a result of the spin-orbit coupling between $m=0$ ($\sigma$) and $m=1$ ($\pi$) states, which vanishes in the atomic limit, both at small and large $R$.

In conclusion, we have introduced a pure exponential basis set, in conjunction with restricted kinetic balance conditions, and shown that it allows for efficient iterative resolution of the Dirac equation for the hydrogen molecular ion. The accuracy of the ground-state energy is improved by about 7 orders of magnitude with respect to previous works. The iterative method furthermore avoids cumbersome algebraic manipulations that are typically required in perturbation theory to regularize divergent expressions. The fact that the energy correction at the fourth iteration is well converged implies that the relativistic wavefunction is accurate up to the third iteration, i.e. at least up to an order of $c^{-6}$. This is an important step towards nonperturbative calculations of the one-loop self-energy correction in hydrogen molecular ions.

\begin{table}[t]
\begin{center}
\begin{tabular}{@{\hspace{3mm}}c@{\hspace{3mm}}l@{\hspace{3mm}}}
\hline\hline
$R$ & $\hspace{2.2cm} E$ \\
\hline      
$0.2$   &  $-1.928\,696\,929\,923\,044\,907\,800(3)$   \\
$0.5$   &  $-1.735\,028\,271\,055\,552\,023\,828\,5(9)$   \\
$1.0$   &  $-1.451\,804\,005\,087\,137\,677\,811(2)$   \\
$2.0$   &  $-1.102\,641\,581\,032\,577\,164\,12(1)$    \\
$3.0$   &  $-0.910\,901\,679\,231\,133\,022\,10(2)$    \\
$5.0$   &  $-0.724\,425\,920\,325\,466\,271\,964(3)$ \\
$7.0$   &  $-0.648\,457\,452\,933\,341\,174\,206\,2(2)$ \\
\hline\hline
\end{tabular}
\end{center}
\caption{Ground-state energies for different values of $R$ obtained using the iterative method.}\label{tab_allrR}
\end{table}

\textbf{Acknowledgements.} H.D.N. and J.P.K. acknowledge support of the French Agence Nationale de la Recherche (ANR) under Grant No. ANR-19-CE30-0029. V.I.K. acknowledges support of the Russian Foundation for Basic Research under Grant No. 19-02-00058-a.


\begin{thebibliography}{99}
\bibitem{DPT-comment} The discrepancy in $E_2^p$ with Ref.~\cite{Korobov07} has been traced back to a slight error in the numerical integration of one of the contributions in that reference.
\end{thebibliography}
\end{document}